\documentclass[sigconf]{acmart}
\AtBeginDocument{%
	\providecommand\BibTeX{{%
			\normalfont B\kern-0.5em{\scshape i\kern-0.25em b}\kern-0.8em\TeX}}}

\setcopyright{acmcopyright}
\copyrightyear{2019}
\acmYear{2019}
\acmDOI{10.1145/1122445.1122456}

\acmConference[KDD '23]{KDD '23: The 29th ACM SIGKDD Conference on Knowledge Discovery and Data Mining}{August 6 - 10, 2023}{Long Beach,CA,USA}
\acmBooktitle{Woodstock '19: ACM Symposium on Neural Gaze Detection,
	June 03--05, 2019, Woodstock, NY}
\acmPrice{15.00}
\acmISBN{978-1-4503-9999-9/18/06}

\usepackage{graphicx}
\usepackage{amsmath}
\usepackage{nicefrac}       
\usepackage{pgfplots}
\usepackage{pgfplotstable}
\pgfplotsset{width=7cm, compat=1.13}

\renewcommand\footnotetextcopyrightpermission[1]{} 
\usepackage{algorithm}  
\usepackage{algpseudocode}  
\usepackage{amsmath}  

\usepackage{multirow}
\usepackage{subfigure}
\usepackage{parskip}
\usepackage{makecell}

\begin{document}

	\title{Audience Expansion for Multi-show Release Based on an Edge-prompted Heterogeneous Graph Network}


 \author{Kai Song}
\authornote{The work was done during their stay in Tencent Video and both authors contributed equally.}
\affiliation{%
  \institution{Tencent Video}
  \city{Beijing}
  \country{China}}
\email{quantumks838@gmail.com}

\author{Shaofeng Wang}\authornotemark[1]
\affiliation{%
  \institution{Tencent Video}
  \city{Beijing}
  \country{China}}
\email{wangsf199610@gmail.com}

\author{Ziwei Xie}\authornotemark[1]
\affiliation{%
  \institution{Tencent Video}
  \city{Shenzhen}
  \country{China}}
\email{zeavidxie@tencent.com}

\author{Shanyu Wang}
\affiliation{%
  \institution{Tencent Video}
  \city{Beijing}
  \country{China}}
\email{sammywang@tencent.com}

\author{Jiahong Li}
\affiliation{%
  \institution{Tencent Video}
  \city{Beijing}
  \country{China}}
\email{kristenli@tencent.com}

\author{Yongqiang Yang}
\affiliation{%
  \institution{Tencent Video}
  \city{Beijing}
  \country{China}}
\email{aronyang@tencent.com}

	
	\begin{abstract}
		
		In the user targeting and expanding of new shows on a video platform, the key point is how their embeddings are generated.  It's supposed to be personalized from the perspective of both users and shows.  Furthermore,  the pursue of both instant (click) and long-time (view time) rewards, and the cold-start problem for new shows bring additional challenges. Such a problem is suitable for processing by heterogeneous graph models, because of the natural graph structure of data. But real-world networks usually have billions of nodes and various types of edges. Few existing methods focus on handling large-scale data and exploiting different types of edges, especially the latter. In this paper, we propose a two-stage audience expansion scheme based on an edge-prompted heterogeneous graph network which can take different double-sided interactions and features into account. In the offline stage, to construct the graph, user IDs and specific side information combinations of the shows are chosen to be the nodes, and click/co-click relations and view time are used to build the edges. Embeddings and clustered user groups are then calculated.  When new shows arrive, their embeddings and subsequent matching users can be produced within a consistent space. In the online stage, posterior data including click/view users are employed as seeds to look for similar users. The results on the public datasets and our billion-scale data demonstrate the accuracy and efficiency of our approach.

	\end{abstract}

\ccsdesc[500]{Computing methodologies~Learning latent representations}
\ccsdesc[500]{Computing methodologies~Neural networks}

	\keywords{user expansion, heterogeneous graph network, look-alike}
	
	
	\maketitle
	\setlength{\parindent}{2em}
	\section{introduction}

For a video platform, high-quality shows are especially important for its prosperity, or in technical words, for its daily active users (DAU) and return of investment (ROI). 

Generally, new shows are released asynchronously with an update speed of some episodes each week. Each show lasts a couple of weeks. Thus, on one certain day (e.g., August 12), more than one show can exist. It's supposed to push these shows in time to users interested in them in our platform. Thus personalized user targeting plays an important role here.

In the calculation of our ROI and other core indices, although click is pursued, view time, one-day retain ratio (the number of users who return the next day after their first login today) and platform interactions  (voting up, comments, etc.) are all important factors. Different from the instant rewards like CTR/CVR in e-commerce or advertising, these longtime rewards are not that well-described in traditional instant reward supervised learning. A click-or-not classifier model or view time regression model are found to bring severe biases for the pursue of the core indices (which are mentioned in Section 5.3 as the base model). And a heavy multi-task model or reinforcement learning frame is a bit of too expensive to tune on our data.


Furthermore, in the release, the side information (categories, directors, actors, plots, etc.) of a new show becomes known to the engineers usually only one day before their online delivery, and thus the show-user interactions are lacked, which is recognized as the new item cold-start problem. In addition, the shows of different types (let's say, romance or war) are usually released during the same period but the click/view variance among them tend to be quite large. Therefore, a CTR model or a 
two-tower model based on historical data combined with transfer learning \cite{dewet2019finding} or meta-learning\cite{vanschoren2018meta} can hardly fulfill our requirements. 

To address the above issues, in this work we make an attempt to use the heterogeneous graph network model. In the graph construction, user IDs and the side information of the shows (instead of the show IDs) are selected as the multi-attribute nodes. Instant and deep long-time rewards between users and shows are taken into account through multi-attribute edges. They are used in the node-node weight calculations and as explicit double-sided features which prompt the model to learn different relationships.

In this way, the embedded vectors of users and combinations of side information of the shows are obtained in the same space. When a new show (and its side information) arrives, its embedded vector can be looked up instantly. When new combinations appear, simple statistical counts or a supervised learning model can help to find their neighboring vectors with good interpretability. In Section 5, we show the high accuracy of this ad-hoc operation for new shows. 

To better describe the interactions between nodes, an edge-prompted heterogeneous graph neural network(EPGNN) is proposed in this work. Explicit interaction terms are added on the edges which should be normalized. They are used to guide the network to learn subgraphs composed of different relationships between nodes. The public dataset evaluation and online results demonstrate the efficiency. Please refer to section 5.1 and 5.3.

Like \cite{dewet2019finding,zhu2021learning}, our framework is also a two-stage scheme. In the offline stage, embeddings of users and shows are generated using the EPGNN model. Then hundreds of millions of user embeddings are clustered into 200,000 groups. The vectors of 200,000 centroids are then obtained. Top N (tens of millions) users of each show are stored as the base part. 
In the online stage, users who click or view a specific show are returned sub-hourly from the online service.  The click/view users of each show are then used as seeds to recall millions of users in sub-minute through the help of the clustered groups and centroids. The details are presented in Section 4.

Our contributions are listed as follows:
\begin{itemize}
	\item Embeddings of users and side information combinations of shows are generated in the same graph space, which could address our cold-start dilemma and be directly used in the release of new shows.
	\item A novel edge-prompted heterogeneous graph neural network (EPGNN) is proposed, which takes subgraphs composed of different relationships between nodes into account. An on-the-fly sampling policy is adopted.
	\item A two-stage (online and offline) scheme is designed with asynchronous update frequency. The offline part is updated daily while the online part is updated sub-hourly.
	\item Clustering methods are applied to accelerate show-user recall and are reused in multiple Dynamic Product Ads (mDPA). 
\end{itemize}
        \section{related work}

\subsection{Audience Expansion}

Look-alike techniques for user expansion are popular for advertisement platform and user growth. They are famous for their lightweight and robustness. During the past decade, various ideas have been proposed. 

Simple rule-based methods\cite{mangalampalli2011feature,shen2015effective}
are naturally the basic start. Data scientists give a couple of user selection policies based on their analyses. For instance, males who are 20-40 year-old and sports fans are selected. In modern views, these can be regarded as single-feature models. Compared to black-boxing networks, interpretability is satisfying while performance  is not bad. In our experiments(Section 5.3), rule-based method is also tested.

In post-deep learning days, it's typical to train a supervised neural network model on historical data, which can then be used to generate embeddings or fine-tuned on new data using transfer learning \cite{perlich2014machine} or meta learning \cite{vanschoren2018meta,zhu2021learning}. In \cite{ma2016score}, Ma et al. introduced a couple of ad-hoc similarity measures in terms of counting and important features. Perlich et al. \cite{perlich2014machine} concentrated on sampling techniques in the application of transfer learning for targeted display advertising. In \cite{dewet2019finding}, deWet et al. from Pinterest trained a supervised classifier model using historical users and their corresponding interacted items (Pins). User vectors were obtained. Transfer learning was applied for updating. In practice, the classifier model and the embedding-based approach were blended. Zhu et al. \cite{zhu2021learning} proposed a similar attack:  meta-learning was utilized to train a general model based on all historical data in the offline stage. Then in the online stage, a customized model evolved from the general model for a new campaign.

\subsection{Graph Neural Network(GNN)}

Deepwalk by Perozzi et al. \cite{perozzi2014deepwalk} and node2vec by Grover et al. \cite{grover2016node2vec} are two famous homogeneous graph embedding models based on word2vec\cite{church2017word2vec}. The former used depth first search (DFS) strategies on the graph to generate sequences while the latter used two parameters $p$ and $q$ to control the superposition of breath first search (BFS) and DFS.

In \cite{dong2017metapath2vec}, the metapath2vec model generalized the random walk to heterogeneous graphs. the meta-path-guided walk strategy was developed, which in principle could capture correlations of multi-attribute nodes and edges.

Velickovic et al. \cite{velickovic2018gat} and Wang et al. \cite{wang2019hgan} leveraged attention mechanism to graph data, which allowed for implicitly assigning different importance to different neighboring nodes in message passing.

Cen et al. \cite{cen2019gatne} proposed a framework to deal with heterogeneous graph with multi-attribute edges. The model was trained under the meta-path scheme.

        \section{framework}

\begin{figure*}[h]
  \centering
  
  \includegraphics[width=\linewidth]
  { 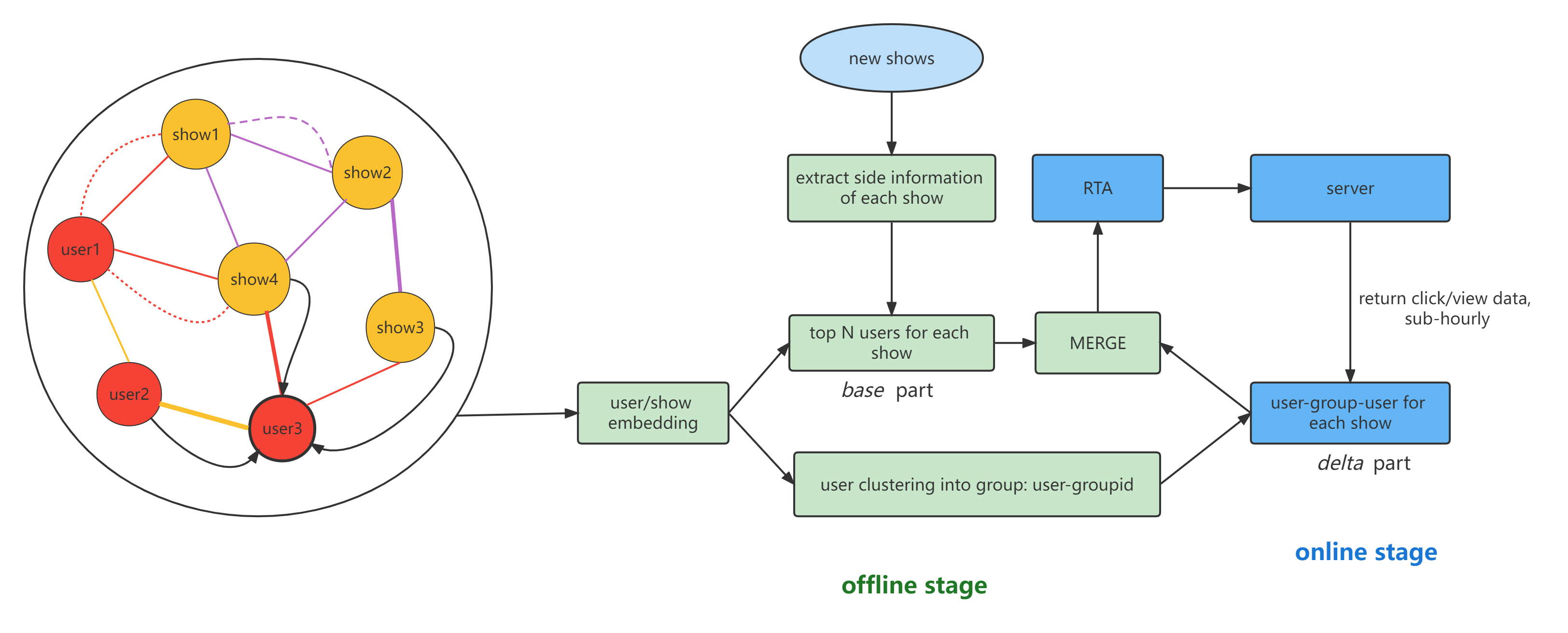}
  \caption{Overall framework}
  \label{pic-framework}
\end{figure*}

Figure \ref{pic-framework} depicts the framework of our system. It's designed to be two-stage similar to a couple of previous works\cite{dewet2019finding,liu2020two}.

In the left circle is a sketch of our heterogeneous graph, which is the workhorse of the nodes' embeddings. The different colors of the nodes and edges mean that they have different attributes. 
The red nodes are the users and the orange nodes are the combinations of the shows' side information. The dashed lines represent view time and the solid lines represent click or co-click relations.
The thickness of the edges means the different importance weight.
The solid black lines depict the message passing to a specific node from its topological environment.
The graph construction details would be described in Section 4.1. In our multi-show release, the graph would be updated daily.

The middle green part is the offline stage. When new shows to be released are on schedule, their profiles (directors, genres, etc.) usually become known one day in advance. Their vectors can be generated by looking up the combination vector space or by simple counts. 
For the former, a supervised regression model can be trained where the shows' profiles are the features and the embeddings are the labels. 
For the latter, We could search for the historical shows whose profiles are overlapped with the new shows'. The embedding vectors of the new shows are then obtained by averaging the embeddings of their similarities. 
In the offline stage, the top N (that's tens of millions) user candidates are calculated for each show. They are used as the $base$ part to serve online. 
On the other hand, all the user embeddings are clustered into 200,000 fine-grained  groups. The bijection between user and groupID is obtained. 

The right blue part is the online stage, which is realized with our in-house real-time API (RTA) service. When the impression of each show starts, the click/view data are returned sub-hourly. These are used as the seeds to expand potential user sets. Actually, embedding distance calculations between user groups and shows are not needed here. Simple counts of their distributions in the 200,000 are precise enough. GroupIDs (and their corresponding users) which contain the most seeds are selected. This can also help to alleviate the noise in seeds. As the $delta$ part, they are merged with the $base$ part to serve online.
        \section{heterogeneous interactive graph neural network}

\subsection{Graph Construction}
For new shows, there are no click/view data before they are released. Thus they cannot be included in the graph if we use the show IDs as nodes directly. The combinations of their side information are used instead, and the generality to new shows of our graph model becomes intrinsic, which helps to alleviate the cold-start problem in our case.

About 25 kinds of side information of the shows are selected, which includes main genre, sub genre, directors, writers, leading actors, rating, producers, hot levels, etc. We have to point that the distinct numbers of these profiles are large and their combinations are incredible. Luckily, in practice most of the combinations are meaningless (for example, some actors' co-operations with most foreign directors  are rare events). About 1.35 millions of profile combinations are finally chosen. 

Other fingerprint profiles are not selected to suppress noises.

\begin{table*}[]
\caption{Data details}
\label{tab:data-description}
\begin{tabular}{|c|cc|cc|c|}
\hline
\multirow{2}{*}{\textbf{Description}} & \multicolumn{2}{c|}{\textbf{node}}                   & \multicolumn{2}{c|}{\textbf{homogenenous edge}}            & \textbf{heterogeneous edge} \\ \cline{2-6} 
                                      & \multicolumn{1}{c|}{show}             & user        & \multicolumn{1}{c|}{show-show}      & user-user          & show-user                  \\ \hline
calculation logic                                 & \multicolumn{1}{c|}{show combination} & userid      & \multicolumn{1}{c|}{same type or not} & click the same show & click, view time             \\ \hline
Number                                & \multicolumn{1}{c|}{1.35+ million}     & 300 million & \multicolumn{1}{c|}{200 million}       & 1 billion        & 3 billion                   \\ \hline
\end{tabular}
\end{table*}

Table \ref{tab:data-description} gives the node and edge details of our big graph and Table \ref{tab-degree} shows the degree distribution of each kind of edges. We could tell that some shows are outstandingly hot.
\begin{table}[]
\caption{Degree distribution}
\label{tab-degree}
\begin{tabular}{c|c|c|c}
\textbf{edge type} & \textbf{quantile-1}              & \textbf{quantile-2}              & \textbf{quantile-3}              \\ \hline
show-show       & \textgreater{}100 2.0\% & \textgreater{}20 15.6\% & \textgreater{}10 29.9\% \\
user-show       & \textgreater{}40 7.2\%  & \textgreater{}20 16.7\% & \textgreater{}10 33.8\% \\
user-user       & \textgreater{}20 44.1\% & \textgreater{}10 53.8\% & \textgreater{}5 63.8\% 
\end{tabular}
\end{table}

It should be mentioned that, more kinds of nodes including the actors and directors of the shows can of course be included in the graph. And, if the billion-scale users are too heavy, they can be firstly clustered into course-grained groups based on their features or other priory information. These user groups are used instead to build the graphs with the combinations of the shows. The graph becomes only million-scale and can be handled quite easily.

\subsection{Edge-Prompted Graph Neural Network (EPGNN)}

Graph network provides an elegant frame to consider multiplex interactive elements (nodes) in a consistent space elegantly. Heterogeneous graph network refers to the general case where graphs consist of two or more kinds of nodes or edges, or both.

The paradigm of passing messages on graphs can be roughly divided as graph embedding and graph network. The former includes classic models such as DeepWalk\cite{perozzi2014deepwalk}, node2vec\cite{grover2016node2vec}, EGES\cite{wang2018billion}, etc. The latter includes the seminal GCN\cite{kipf2016semi}, graphSAGE\cite{hamilton2017graphsage}, GAT\cite{velickovic2018gat}, etcs. Graph network can further be divided into spatial and spectral types, depending on whether integral transformation (Fourier, Laplace or others) is performed to change spatial relationships into spectral domains. GraphSAGE belongs to the spatial type while GCN belongs to the spectral method.

In this paper, we present an Edge-Prompted Graph Neural Network (EPGNN) model which is applicable to graphs with multi-attribute nodes and multi-attribute edges.
EPGNN can make full use of the different relationships between heterogeneous graph nodes, and perceive subgraphs composed of any relationship. Like graphSAGE, it is of spatial type.

\begin{figure*}[h]
  \centering
  \includegraphics[width=0.8\linewidth]
  {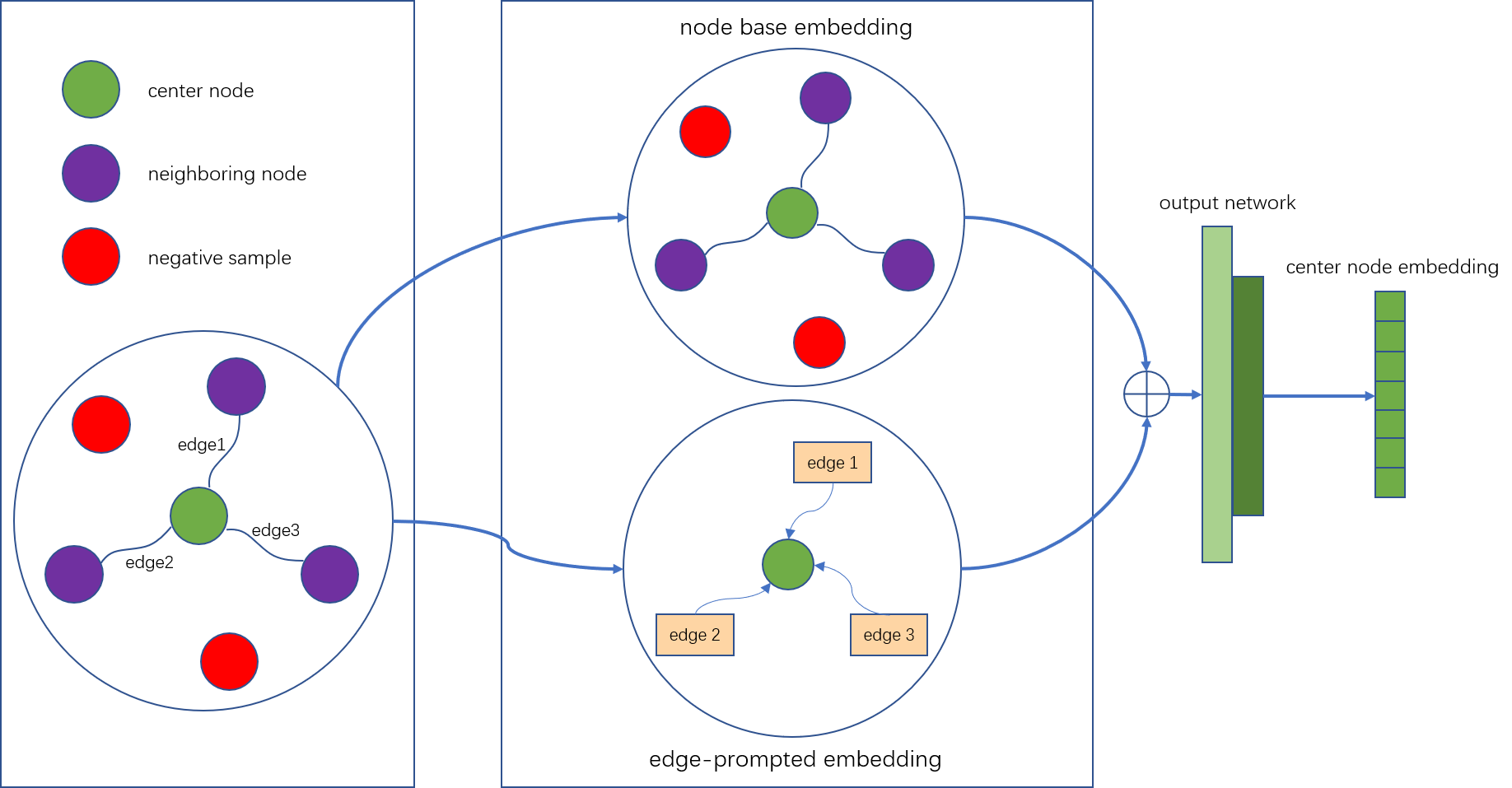}
  \caption{The model structure of the proposed EPGNN}
  \label{pic-model}
\end{figure*}

Inspired by the message passing neural network (MPNN)\cite{gilmer2017neural}, our iteration is designed as follows:

The embedding of node $i$ at iteration $t+1$ is:
\begin{equation}
v^{t+1}_{i} = b^t_i+f(v^t_i)+g(\{u^t_{ij}\},\{v^t_j\},v^t_i)
\end{equation}
 

Here, $v^{t+1}_{i}$ is of shape $m_n \times 1$ and can be initialized using the side information (features) of this node.

$u^t_{ij}$ is the embedding of edge with shape $m_e\times 1$, which does not have to be the same dimension as nodes'.

 $b_i$ is the bias term of shape  $m_n \times 1$, randomly initialized.

$f(v^t_i)$ is the one-body term that processes single-sided information of node $i$ and would be designated as a multi-layer neural network without loss of generality.

$g(\{u^t_{ij}\},\{v^t_j\},v^t_i)$ is the two-body interactive term which consider the different interactions between node $i$ and its neighboring nodes (or neighbors of neighbors, of course): $v_j \in\mathcal{N}(v_i)$. Weighting the features of neighboring nodes through double-sided information, the network can learn subgraphs constructed from various relationships without interfering with each other. The attention technique can be integrated:
\begin{equation}
g(\{u^t_{ij}\},\{v^t_j\},v^t_i) = \sum_j \alpha_{ij}(v^t_i,v^t_j)(\sum_k u^t_{ijk} W_{ jk}v^t_j),
\end{equation}
where $\alpha_{ij}$ is the attention coefficient (a real number) and could be defined using a single  feedforward layer\cite{velickovic2018gat}:
\begin{equation}
\alpha_{ij} = \sigma\big(F_{\rm nn}[v^t_i,v^t_j]\big)\;,
\end{equation}
with $\sigma$ being a nonlinear mapping (sigmoid function for instance) and $F$ representing the transformation parameters.

$W_{jk}$ is a trainable node-node matrix of shape $m_n\times m_n$.

$u^t_{ijk}$ is an an element in the vector $u^t_{ij}$. Its value represents a certain double-sided interaction such as click or view time.



Eq.(1) is pretty general. If we keep only the first two terms, it is just a deep neural network. If we keep the three terms with/without side information, setting $u^t_{ij}$ to be a constant, it is roughly equivalent to GraphSAGE/node2vec. If we keep the three terms without double-sided features, setting $u^t_{ij}$ to be a scalar, it degenerates into GATNE\cite{cen2019gatne}.


The  log-sigmoid loss function is adopted \cite{church2017word2vec,hamilton2017graphsage}:
\begin{equation}
\mathcal L(v_i) = -\log (\sigma(v_k^Tv_i)) - Q \mathbb {E}_{v_j \in P_{\rm neg}(v)} \log(\sigma(-v_j^Tv_i))
\end{equation}
where the first term measures the similarity of neighboring nodes and the second term measures the distance between a node and its negative samples in the graph.

Here $\sigma$ is the sigmoid function.  $P_{\rm neg}(v)$ is a negative sampling distribution.  $Q$ is a real number that controls the influence of negative terms.

Our EPGNN model can be adaptable to both transductive learning and inductive learning. 
The model, in essence, learns the mapping between graph information and node embedding in an unsupervised way. 
Unlike models like node2vec, we don't have to include certain nodes in the training to generate 
their embeddings.
The process of single-sided features borne by nodes, double-sided features borne by edges and topological information can be  decoupled from the training part as shown in Algorithm 1. And, in the large-scale graph scenario, all these can be done in parallel. 

\noindent\textbf{Policy of Hierarchical Neighbor Sampling}

To compute the third term in Eq.(1), a node and its neighboring nodes/edges (or neighbors of neighbors, etc.) are needed.
To deal with the constraints of computing resources (CPU and especially memory) and the challenge of expanding to large billion-node graphs, we design a hierarchical  sampling  policy. 
In the first-order sampling, about 70\% (or more, as long as the memory allows) of a node's neighboring nodes and edges are randomly selected to store into the memory. 

The second-order sampling proceeds in an on-the-fly way. For each batch, 50\% of 
the stored 70\% neighboring nodes and edges are re-sampled (that's 35\% of the original degrees) with replacement.
To this end, the speed of the iteration  is guaranteed while memory-allowable information from a node's locality is not lost.

Yet, the random selection of neighboring nodes/edges can be improved by virtue of the attributes and degree distributions (Table 2) of the nodes/edges.


\noindent\textbf{Forward Propagation Algorithm for EPGNN}

As shown in Algorithm 1, our forward propagation algorithm consists of four sections: graph construction, sequence generation, multi-hop pair generation and model training. 

In lines 6-9, monthly active user and show as nodes are selected to construct $G_1$ in the tuple form of (source, destination, weight). The weights are calculated in the way as explained in Section 4.1. 

Lines 10-15 correspond to the sequence generation part, where the metapath2vec\cite{dong2017metapath2vec} algorithm is employed.

Lines 16-17 generate positive samples by window sliding on the sequences.
Compared with the original graph data $G_1$, multi-hop relations are introduced instead of only one-hop neighboring ones. The sliding window size could control this long range interactions.

Lines 18-21 are the decoupled preparation for features of neighboring nodes and edges.

Lines 22-27 are the EPGNN training part.

Line 19 and 24 are our hierarchical neighboring sampling policy as explained above.


This billion-scale large graph is a great challenge to storage and computing. To keep the computation cost under control and to accelerate the meta2path, we turn to the use of aprioir graph partition based on our business context. To be specific, domain knowledge and a series of data analyses tell that the interactions among the shows of different first class category are relatively sparse. The original graph is cut along these edges into ten parts, which are then fed into SequenceGeneration in parallel.

\begin{algorithm}[htb]  
  \caption{Forward propagation algorithm for EPGNN}  
  \label{alg:Framwork}  
  \begin{algorithmic}[1]  
    \Require  
      \\
      the set of nodes $V$; 
      the set of node features $F_n$; 
      the set of edge features $F_e$;  \\
      meta-paths;
      window size $k$;
      walk length $l$;
      walks per node $r$; \\
      first-order sample number $s_1$;
      second-order sample number $s_2$; \\
      EPGNN training epoch $n$;\\
      $walks= []$ 
    \Ensure  
       embeddings $\{v\}$ for all v $\in V$;  
    \State edge set $E_{\rm user-show}=\{(\rm v_{\small userID},\rm v_{\tiny show})\}$
    \State edge set $E_{\rm user-user}=\{(\rm v_{userID},\rm v_{\small userID})\}$
    \State edge set $E_{\rm show-show}=\{(\rm v_{\tiny show},\rm v_{\tiny show})\}$
    \State $G_1(V,E_1)=\textrm {GraphGeneration}(E_{\rm user-show},E_{\rm user-user},E_{\rm show-show})$, in the form of $\rm (v_{source}, v_{destination}, weight)$
    \For{$iter=1$ to $r$}
      \For{all nodes v $\in V$}
        \State $walk = \textrm {SequenceGeneration}(G_1,{\rm v},l)$
        \State Append $ walk$ to $ walks$
      \EndFor
    \EndFor
    \State $pair\_set = \textrm {PairGeneration}(walks,k)$
    \State $G_2(V,E_2) = \textrm{ SampleGenerationFromPair}(pair\_set, G_1(V,E_1))$
    \For {all nodes v $\in V$}
      \State     
      first-order sampling,
      sample $s_1$ neighbor nodes $\{neighbor\}_{\rm v}$ of $v$ with corresponding node features $\{f_n\} \subseteq F_n$ and edge features $\{f_e\} \subseteq F_e$
      \State $\textit{neighbor}\_info_{\rm v} = \{\textit{neighbor},f_n,f_e\}_{\rm v}$
    \EndFor
    \For{ $iter=1$ to $n$}
      \For{$\rm (v_i,v_j)$ in $pair\_set$}
        \State
        second-order sampling,
        $\{neighbor^\prime,f_{n}^\prime,f_{e}^\prime\}_{\rm v_i}\leftarrow s_2$ items from $neighbor\_info_{\rm v_i}$
        \State Update $v_{\rm v_i}$ by $Adam({\rm v_i},{\rm v_j},\{\textit{neighbor}^\prime,f_{n}^\prime,f_{e}^\prime\}_{\rm v_i})$
      
      \EndFor
    \EndFor
    \State
    \Return $\{v\}$;  
  \end{algorithmic}  
\end{algorithm}

        \section{experiments}

\subsection{Performance Analysis}

\begin{table*}[htbp]
\centering
\caption{Performance comparison of different methods on three datasets}
\begin{tabular}{c|ccc|ccc|ccc}
\hline
             & \multicolumn{3}{c|}{Amazon}                      & \multicolumn{3}{c|}{YouTube}                     & \multicolumn{3}{c}{Twitter}                      \\
             & ROC-AUC        & PR-AUC         & F1             & ROC-AUC        & PR-AUC         & F1             & ROC-AUC        & PR-AUC         & F1             \\ \hline
DeepWalk     & 94.20          & 94.03          & 87.38          & 71.11          & 70.04          & 65.52          & 69.42          & 72.58          & 62.68          \\
node2vec     & 94.47          & 94.30          & 87.88          & 71.21          & 70.32          & 65.36          & 69.90          & 73.04          & 63.12          \\
LINE         & 81.45          & 74.97          & 76.35          & 64.24          & 63.25          & 62.35          & 62.29          & 60.88          & 58.18          \\ \hline
metapath2vec & 94.15          & 94.01          & 87.48          & 70.98          & 70.02          & 65.34          & 69.35          & 72.61          & 62.70          \\ \hline
PMNE         & 88.38          & 88.56          & 79.67          & 70.61          & 69.82          & 65.39          & 62.91          & 67.85          & 56.13          \\
MVE          & 92.98          & 93.05          & 87.80          & 70.39          & 70.10          & 65.10          & 72.62          & 73.47          & 67.04          \\
GATNE      & \textbf{97.39} & \textbf{97.01} & 92.62          & 84.36          & 82.06          & 76.95          & 74.96          & 74.93          & 68.29          \\ \hline
EPGNN(ours)   & 97.36          & 96.68          & \textbf{92.93} & \textbf{85.00} & \textbf{82.73} & \textbf{77.28} & \textbf{81.02} & \textbf{80.11} & \textbf{73.18} \\ \hline
\end{tabular}
\end{table*}

Link prediction is widely used to evaluate the quality of the node embeddings. So we first compare our model with other state-of-the-art algorithms on three public datasets. Amazon Product Dataset\footnote{http://jmcauley.ucsd.edu/data/amazon/} \cite{he2016ups,mcauley2015image} includes product metadata and links between products; YouTube dataset\footnote{http://datasets.syr.edu/datasets/YouTube.html} \cite{tang2009uncovering,tang2009uncoverning} consists of various types of interactions; Twitter dataset\footnote{https://snap.stanford.edu/data/higgs-twitter.html} \cite{de2013anatomy} also contains various types of links.

Three types of algorithms are selected to be the competitor: network embedding methods, heterogeneous network embedding methods, and multiplex heterogeneous network embedding methods. The network embedding methods include DeepWalk, LINE\cite{tang2015line}, and node2vec, which can only deal with homogeneous graph. for heterogeneous network embedding methods, we focus on the representative work metapath2vec. And we select PMNE\cite{liu2017principled}, MVE\cite{qu2017attention}, GATNE as representatives of the multiplex heterogeneous network embedding methods, in which the GATNE is our baseline. We reproduced the experimental results of GATNE, and ensured that the training settings of our EPGNN model are consistent with it.

The experimental results of three public datasets are shown in Table 3. Our model outperforms all types of baseline models in F1-score on various datasets. As we can see, All indicators of our method are the best on YouTube and Twitter dataset. Especially on the Twitter dataset, our model outperforms the baseline model GATNE by about six percentage points.

\subsection{Posterior Analyses and Offline Evaluation}

\noindent\textbf{Visualization of the Generated Embeddings}

\begin{figure}[htbp]
    \centering
    \subfigure[]{
        \includegraphics[width=2.5in]{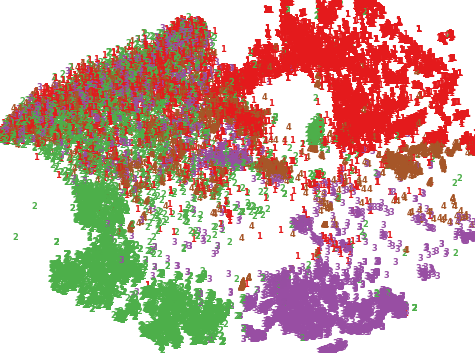}
    }
    \quad    
    \subfigure[]{
    	\includegraphics[width=2.5in]{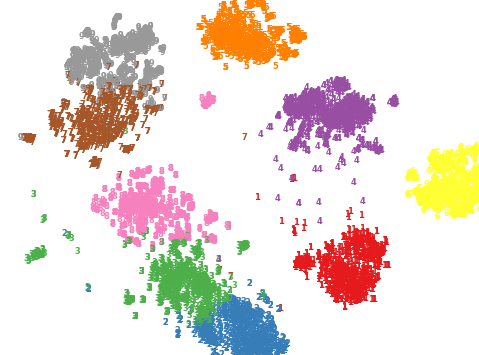}
    }
    \caption{Visualization of the show vector space by reducing the dimension from 32 to 2 using t-SNE}
    \label{pic-cidembed}
\end{figure}

Figure \ref{pic-cidembed} is the visualization of the show vector space by reducing the dimension from 32 to 2 using t-SNE\cite{van2008visualizing} algorithm for the shows. The same color in Figure \ref{pic-cidembed} represents the show combinations from the same category/sub-category.
In the upper plot, 5 different course-grained categories are depicted. We could see that
the separation is relatively good, which proves the quality of our generated latent vectors. However, the up right corner is kind of fuzzy. We suspect the categories do have superposition (for instance, some films are both "films" and "documentaries" ).  To test this analysis, 9 finer categories obtained by considering the types of the shows are presented, as shown in the lower part of Figure  \ref{pic-cidembed}.
Now the separation boundaries become much more sharp.


\noindent\textbf{Uplifts of Online Hourly-Updating Part over Offline Daily-Updating Part}

\begin{table}[h!]
  \begin{center}
    \caption{CTR and effective view uplift results of $delta$ part over $base$ part}
    \label{tab-ctrcvr}
    \begin{tabular}{c|c|c} 
      \textbf{Hour} & \textbf{CTR uplift} & \textbf{effective view uplift}\\
      \hline
      16:00 & 5.92\% & 4.14\%\\
      17:00 & 7.70\% & 3.57\%\\
      18:00 & 3.42\% & 9.09\%\\
      19:00 & 3.44\% & 4.99\%\\
      20:00 & 10.31\% & 16.86\%\\
      21:00 & 3.88\% & 10.87\%\\
      22:00 & 2.16\% & 4.96\%\\
      23:00 & 3.83\% & 9.16\%\\
    \end{tabular}
    
  \end{center}
  
\end{table}

\begin{figure}[h]
  \centering
  \includegraphics[width=\linewidth]
  {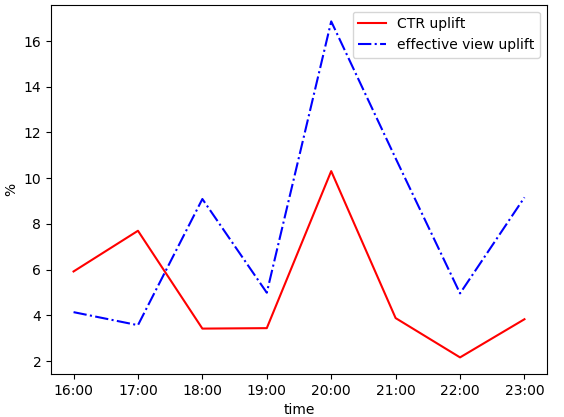}
  \caption{Curves of uplifts of online hourly-updating $delta$ part over offline daily-updating $base$ part }
  \label{graph-ctrcvr}
\end{figure}

Figure \ref{graph-ctrcvr} and Table \ref{tab-ctrcvr} present the contrast of daily-updating $base$ part and hourly-updating $delta$ part. CTR and effective view are both core values in our ROI computation.
We could see that the CTR and effective view of the $delta$ part are much better than the daily-updating $base$ part. CTR is enhanced by an average of 3\% and effective time by 8\%.

In practice, we have optimized the update frequency of $base$ and $delta$ parts respectively and find that the $base$ could be updated every one day and the $delta$ increment should be updated every one hour. One reason for this one-hour frequency is that, the interests and behaviors of the users at 20:00 for example, would possibly be similar to those at 19:00, but relatively different from those at 18:00.

\noindent\textbf{Posterior Analyses and Offline Evaluation}

Here we present the posterior analyses and offline evaluation before online service.

\begin{table*}[h!]
  \begin{center}
    \caption{Offline CTR and effective view results}
    \label{tab-offlineresult}
    \scalebox{1.0}{
   \begin{tabular}{c|c|c|c|c}
\textbf{Time}        & \textbf{\makecell[c]{CTR uplift \\w.r.t. rule based method} }   & \textbf{\makecell[c]{effective view uplift \\w.r.t. rule based method}}  & \textbf{\makecell[c]{CTR uplift \\w.r.t. customized DNN}}   & \textbf{\makecell[c]{effective view uplift \\w.r.t. customized DNN}}   \\ \hline
day1 19:00 & 69.02\%  & 50.16\% & 27.45\%  & 54.27\%  \\
day1 20:00 & 98.52\%  & 57.51\% & 69.86\%  & 81.62\%  \\
day1 21:00 & 109.60\% & 55.34\% & 87.75\%  & 96.53\%  \\
day1 22:00 & 130.12\% & 58.32\% & 104.91\% & 144.25\% \\
day1 23:00 & 24.53\%  & 22.91\% & 13.85\%  & 21.14\%  \\
day2 00:00 & 7.24\%   & 9.49\%  & 5.75\%   & 13.44\%  \\
day2 01:00 & 9.15\%   & 13.90\% & 8.92\%   & 15.91\%  \\
day2 02:00 & 7.86\%   & 10.36\% & 8.06\%   & 11.77\%  \\
day2 03:00 & 4.15\%   & 8.18\%  & 2.76\%   & 9.43\%   \\
day2 04:00 & 4.12\%   & 6.29\%  & 6.78\%   & 9.65\%   \\
day2 05:00 & 8.90\%   & 13.43\% & 7.20\%   & 14.48\%  \\
day2 06:00 & 7.62\%   & 11.84\% & 8.47\%   & 13.38\%  \\
day2 07:00 & 7.20\%   & 9.82\%  & 8.23\%   & 15.29\%  \\
day2 08:00 & 6.75\%   & 11.99\% & 5.09\%   & 15.18\%  \\
day2 09:00 & 38.87\%  & 50.46\% & 14.28\%  & 51.35\% 
\end{tabular}
    }
  \end{center}
\end{table*}

Table \ref{tab-offlineresult} shows comparison CTR and effective view (larger than certain seconds) between different models. 
The rule-based method is similar to references  \cite{mangalampalli2011feature,shen2015effective}. The difference between the customized DNN method and our EPGNN is only the way of generating embedding. 
They adopt the same framework as presented in Section 3.

we can see that our model EPGNN could significantly improve both the CTR and effective view.

\subsection{Online Evaluation}

\begin{table}[h!]
  \begin{center}
    \caption{Online results of different models}
    \label{tab-onlineresult}
    \scalebox{0.9}{
    \begin{tabular}{c|c|c|c} 
      \textbf{Model} & \textbf{ROI uplift} & \textbf{DAU uplift} &\textbf{ARPU uplift}\\
      \hline
      rule-based & 11.43\% & 26.81\% & -9.26\% \\
     customized DNN & 19.76\% & \textbf{28.18}\% & -3.60\% \\
      EPGNN  & \textbf{28.57\%} & 27.34\% & \textbf{3.00\%} \\
    \end{tabular}
    }
  \end{center}
\end{table}


In Table \ref{tab-onlineresult}, the online results of different methods are compared. 

The base method is chosen to be a general (not customized for show release) CTR model using XGBoost. Three key metrics, ROI, DAU and ARPU (Average Revenue Per User) are listed. We could see that, the rule-based method obtains good ROI and DAU improvements. But the ARPU is contaminated too much.
The ROI and DAU of the customized DNN method are even better than the rule-based model's. And the ARPU decreases less.

 Contrasted with the former two approaches, the EPGNN proposed in this paper has the best boost in ROI, which is more than twice the rule-based model's. Our model's DAU is on the same level as the other two models, and its ARPU does not decrease.

\subsection{Validity of Clustering Embedding}

More than 300 million users are clustered using Faiss\cite{johnson2019billion} and 200, 000 groups are generated. The validity of item-group-user recall scheme is shown in Table \ref{tab-onlinerecall}. We can see that for the four shows, the recall results of item-group-user can cover at least 91\% of the precise item-user way. Therefore, the acceleration brought by vector clustering allows us to deal with much more items. In fact, we have used this approach in the targeting of mDPA (multiple Dynamic Product Ads, which are on the order of 10, 000 in our case) and obtained satisfying results.

\begin{table}[h!]
  \begin{center}
    \caption{Comparison between the results of user-user and user-group-user recalls}
    \label{tab-onlinerecall}
    \scalebox{0.78}{
    \begin{tabular}{c|c|c|c|c} 
      \textbf{show} & \textbf{show-user recall} & \textbf{show-group-user recall} &\textbf{intersection} &\textbf{precision}\\
      \hline
       show1 & 59,945,504 & 62,244,815 & 55,558,198 & 98.1\%\\  
       show2 & 67,877,920 & 71,002,776 & 61,884,203 & 91.2\% \\ 
       show3 & 60,866,837 & 62,675,794 & 56,886,038 & 93.5\% \\ 
        show4 & 70,312,486 & 72,610,760 & 68,409,003 & 97.3\% \\ 
    \end{tabular}
    }
  \end{center}
\end{table}

        \section{Conclusions}

In this work we give an industrial-scale solution to user expansion in multi-show release for a video platform. 
An edge-prompted heterogeneous graph neural network model is proposed.
To take into account of both the instant and the long-time rewards, click/co-click and view time are used to build the edges. To handle the cold-start problem in the release of new shows, the side information combination of a show instead of its ID is chosen to be one kind of the nodes.

The scheme is designed to be two-stage. 
In the offline stage, the graph consisted of users and side information combinations of shows with multi-attribute edges are daily updated and trained to generate node embeddings in the same space. Through the distance computation  between users and each to-be-released show, the $base$ user sets of all shows are obtained. Meanwhile, the user embeddings are clustered into fine-grained groups and the embedding of each group centroid is calculated. 

In the online stage, click/view data of each show are sub-hourly returned from the online service. These click/view users serve as the seed set and help to look for more potential audience (as the $delta$ set) using user-group-user technique. The $base$ and $delta$ parts are merged to serve online.


In our future work, one attempt is the application of dynamic graph combining with meta leaning which could accelerate the updating using hourly incremented data. This is designed for high-frequency activities in the video platform which usually last only a couple of hours and vary dramatically. There, heavy weapons like slowly updated graph or multi-task models can hardly be applied. 
A second direction is the diversity control on the graph. Multiple shows are released at the same time. If the impression largely inclines only to one of them, the other shows cannot accumulate audience. The idea of DPP (Determinantal Point Process) \cite{chen2018fast} can be integrated to search for balance.
And, finally, for new arrivers of our video platform, their information is pretty sparse. Their local environments in the graph are thus rough: too less edges for message passing. We have tested that Betti numbers\cite{cang2017analysis} are helpful tools.

\section{Acknowledgements}

We appreciate the help and discussions of Dong Cui, Jinlong Ran and Xiao Xiong from Tencent Video Growth team. They took an active part in the formalization of Multi-show release. The online service was realized by the RTA (real-time API) of our team.
 We thank Teng Hu for his help on RTA service. We also thank Dr. Lei Fan for his encouragement to publish this piece of work.


	\newpage
	
	\bibliographystyle{ACM-Reference-Format}
	\bibliography{reference}


\end{document}